\documentstyle[11pt,epsfig]{article}
\begin{document}

\thispagestyle{empty}
\begin{flushright}
{CERN-TH/00-000}\\
{FTUAM-00-00}\\
{IFT-UAM/CSIC-00-00}\\
{FTUV-00-09}\\
{IFIC/CSIC-00-09}
\end{flushright}
\vspace*{1cm}  
\begin{center}
{\large{\bf The atmospheric neutrino anomaly without maximal mixing?} }\\
\vspace{.5cm}
A. De
  R\'ujula$^{\rm a,}$\footnote{derujula@nxth21.cern.ch},
M.B. Gavela$^{\rm b,}$\footnote{gavela@garuda.ft.uam.es,}
and 
P. Hern\'andez$^{\rm a,}$\footnote{pilar.hernandez@cern.ch. On leave 
from Dept. de F\'{\i}sica Te\'orica, Universidad de Valencia.}

\vspace*{1cm}
$^{\rm a}$ Theory Division, CERN,
  1211 Geneva 23, Switzerland\\
$^{\rm b}$ Dept. de F\'{\i}sica Te\'orica, Univ. Aut\'onoma de
Madrid, Spain
\end{center}
\vspace{.3cm}
\begin{abstract}
\noindent
We consider a pattern of neutrino masses in which there is
an approximate mass degeneracy between the two mass
eigenstates most coupled to the $\nu_\mu$ and $\nu_\tau$ flavour eigenstates. 
Earth-matter effects can lift this degeneracy and
induce an effectively maximal mixing between these 
two generations. This occurs if $\nu_e$'s 
contain comparable admixtures of the degenerate eigenstates, even
rather small ones.
This provides an explanation of the atmospheric neutrino anomaly in
which the {\it ab initio} introduction of a large mixing angle
is not required. To test this possibility we perform a novel
and detailed analysis of the 52 kiloton-year
SuperKamiokande data, and we find that in a large region of parameter 
space the corresponding confidence levels are excellent. The most recent
results from the Chooz reactor experiment,
however, severely curtail this region, so that the conventional
scenario with nearly maximal mixing angles --which we also
analyse in detail-- is supported by the data.

\end{abstract}

\vspace{2.2cm}
\begin{flushleft}
{CERN-TH/00-000}\\
{FTUAM-00-00}\\
{IFT-UAM/CSIC-00-00}\\
{FTUV-00-09}\\
{IFIC/CSIC-00-09}\\
January 2000
\end{flushleft}
\newpage

\section{Introduction }
\bigskip

The results of the SuperKamiokande collaboration (SK) on the atmospheric 
neutrino deficit \cite{Superka} can be  explained in terms of neutrino 
oscillations \cite{osc}. 
It is natural to analyse the data in the context of the most general 
mixing pattern of three neutrinos, since that is their known number.
Three generations are necessary if
oscillations are to explain the atmospheric and solar~\cite{sol} anomalies: a 
scheme with only two neutrinos cannot account for both effects. 

Let $\bar U$, with 
$(\nu_e,\nu_\mu,\nu_\tau)^T=\bar U\cdot (\nu_1,\nu_2,\nu_3)^T$, be the  
Cabibbo-Kobayashi-Maskawa (CKM) matrix in its most conventional
parametrization, reviewed by the Particle Data Group~\cite{PDG}:
\begin{equation}
\bar U \equiv \bar U_{23} \bar U_{13} \bar U_{12}\equiv
\left(\matrix{
1 & 0 & 0 \cr 
0 & \bar c_{23} & \bar s_{23}   \cr
0 & -\bar s_{23}  & \bar c_{23} \cr}
\right)\;
\left(\matrix{
\bar c_{13} & 0 & \bar s_{13}\, e^{i \delta}\cr
0 & 1 & 0 \cr
- \bar s_{13}\, e^{-i \delta} & 0 & \bar c_{13} \cr}
\right) \;
\left(\matrix{
\bar c_{12} & \bar s_{12} & 0 \cr
-\bar s_{12} & \bar c_{12} & 0 \cr
0 & 0 & 1 \cr}
\right) 
\label{CKM}
 \end{equation}
with $\bar s_{12}\equiv \sin \bar\theta_{12}$, and similarly for the other
sines and cosines. 
Several groups have performed analyses of atmospheric and solar
data in terms of three-neutrino mixing \cite{three}, 
as described by Eq.~(\ref{CKM}), or including sterile
neutrinos~\cite{two}. It is
common to these studies to restrict to a ``minimal scheme'',
in which the mass square difference relevant to  
atmospheric oscillations dominates over the one relevant to 
solar neutrinos: $\Delta \bar m^2_{23}\!\gg\!\Delta\bar m^2_{12}$. 
In this scenario,  
the number of  parameters  describing oscillations at
terrestrial distances is reduced to three: $\bar s_{13}$, 
$\bar s_{23}$ and $\Delta \bar m^2_{23}$, while those 
most relevant to solar
neutrinos are: $\bar s_{12}$, $\bar s_{13}$ and 
$\Delta \bar m^2_{12}$. The best fit of the atmospheric 
data \cite{fogli} is:
\begin{eqnarray}
\Delta \bar m^2_{23} \sim 2.8\times 10^{-3}\, {\rm eV}^2,\;\;\;\; \sin^2 
(2\bar\theta_{23}) \sim 1, \;\;\;\; \bar s^2_{13} \sim 2\times 10^{-2}.
\label{results}
\end{eqnarray}
The angle $\bar \theta_{23}\simeq \pi/4$ is close to maximal, to explain the 
dearth of muons in SK.

The situation for solar neutrino oscillations is less definite~\cite{solanal}. The
combined solar experiments allow for three different 
regions of parameter space. The solar deficit can be interpreted either 
as MSW (matter enhanced) oscillations~\cite{MSW} with an 
angle $\bar\theta_{12}$ that can be large or small, or as nearly-maximal vacuum oscillations, $\bar\theta_{12}\simeq \pi/4$. The corresponding
 mass differences --$\Delta \bar m^2_{12}=10^{-6}$ to $10^{-4}$ eV$^2$, or 
some $ 10^{-10}$ 
eV$^2$-- are significantly below the range deduced from
atmospheric observations, giving support to the minimal
scheme. If $\Delta \bar m^2_{12}\!\sim\!{ O}(10^{-3})$ 
eV$^2$, it can 
have non-negligible effects on atmospheric data that 
have been recently studied~\cite{solaratmos}.

As is well-known, the field of neutrino oscillations is permeated by
a tally of implausible facts and coincidences.
Oscillations over a distance $L$  occur  if
$\Delta m^2 L/ E_\nu\sim 1$. In the various data samples used by
the SK collaboration, the average neutrino
energies are roughly 1, 10 and 100 GeV,
so that for the value of $\Delta m^2_{23}$ in Eq.~(\ref{results}), the ``right''
distances to measure an effect are 280, 2800 and 28000 km: the size of
our planet and the energies in the cosmic ray spectrum  have
been chosen snugly. Something entirely similar can be said
about the low-mass or ``just-so'' solution
to the solar neutrino problem. Moreover, the solar and atmospheric
neutrino ``anomalies'' could have been observed only if the effects are
large. This requires surprisingly large mixing angles, 
except for the ``small-angle''
MSW solution to the solar neutrino problem, for which the ambient matter 
elegantly enhances the effect of a small vacuum mixing. 
Considerable theoretical effort has
been invested in arguing that large neutrino mixings are natural,
as small quark mixings are believed to be.

A final peculiarity of the observed atmospheric neutrino oscillations
involves  the matter contribution to the effective
squared mass of electron neutrinos: 
\begin{equation}
A\equiv 2~\sqrt{2}~G_F~E_\nu~n_e\, ,  
\label{Amat}
\end{equation}
where $n_e$ is the electron number density in the Earth.
 For a typical average terrestrial
density of 5 g/cm$^{3}$ and $E_\nu=10$ GeV, $A\sim 3.7\times 10^{-3}$ eV$^2$,
again in the ballpark of Eq.~(\ref{results}). This last coincidence
suggests the existence of a ``small-angle solution'' to the atmospheric
neutrino problem. 

In the ``small-angle'' scheme we study, the large observed
$\nu_\mu$ disappearance is generated 
in the following way. Let 
$\nu_2$ and $\nu_3$  be almost degenerate, and heavier
than $\nu_1$. When
neutrinos traverse the Earth, the degeneracy is lifted
by matter effects, enhancing
$\nu_\mu$--$\nu_\tau$ oscillations. 
The consequent transitions are maximal if the electron neutrino has
comparable vacuum admixtures of the degenerate mass eigenstates,
$\bar U(e2)\simeq \bar U(e3)$, even if these quantities are not large.

We find that an explanation of  the atmospheric neutrino anomaly
not involving nearly-maximal neutrino
mixing indeed exists, but is
disfavoured by the complementary information from reactor neutrinos.
For the current experimental situation,
large neutrino mixing angles seem to be unavoidable.

The titles of the chapters and appendices specify the
structure of this paper.

\section{Apparent large mixings induced by matter effects}
\subsection{From small to large angles}
In a three-family scenario,
let one neutrino mass eigenstate be much lighter than the
other nearly-degenerate two. Their squared mass matrix
can be written as:
\begin{equation}
M^{vac}_K\simeq\,{\rm Diag}\,\left[\,\mu^2_1;\;m^2 + \mu^2_2;
\; m^2+ \mu^2_3 \,\right] \, ,
\label{m}
\end{equation}
where $K$ stands for the ``kinetic'' eigenbasis (as opposed to the flavour
basis), $vac$ is for vacuum, and $\mu^2_i\!\ll\! m^2$. The degree of degeneracy
assumed for the two heavier neutrinos embodies three conditions: 
their mass difference $\Delta m^2_{23}\equiv\mu^2_3-\mu^2_2$ is
much smaller than their common mass scale $m^2$; it is also 
small enough not
to induce observable oscillations over terrestrial distances 
($\Delta m^2_{23} L/ E_\nu\!\ll\! 1$ for the relevant energies and lengths
of travel); and it is smaller than the effective mass excess
induced on electron neutrinos by matter effects, $\Delta m^2_{23} \!\ll\! A$,
with $A$ as in Eq.~(\ref{Amat}).
{\it De facto}, these conditions simply amount to $\Delta m^2_{23} \!\ll\! 10^{-4}\,{\rm eV}^2$. In practice we can set $\mu_i^2=0$ in the analysis
of oscillations over terrestrial baselines.

For the mass pattern of  Eq.~(\ref{m}), one of the three neutrino
mixing angles and the CP-violating phase  of the CKM
matrix $U$ are unobservable. This is readily checked.
Parametrize the CKM matrix in the unconventional order 
$U \equiv U_{12} U_{13} U_{23}$, as follows:
\begin{equation}
U \equiv U_{12} U_{13} U_{23}\equiv
\left(\matrix{
c_{12} & s_{12} & 0 \cr
-s_{12} & c_{12} & 0 \cr
0 & 0 & 1 \cr}
\right) \; 
\left(\matrix{
c_{13} & 0 & s_{13} \cr
0 & 1 & 0 \cr
-s_{13} & 0 & c_{13} \cr}
\right) \; 
\left(\matrix{
1 & 0 & 0 \cr 
0 & c_{23} & s_{23} \,e^{i \delta}  \cr
0 & -s_{23} \,e^{-i \delta} & c_{23} \cr}
\right)
\label{ourCKM}
\end{equation}
with $s_{12}=\sin(\theta_{12})$, etc.  
The vacuum mass matrix in flavour
space $M^{vac}_F  \equiv U M^{vac}_K U^\dagger$
does not depend on $s_{23}$, or on $\delta$. 
The mixing matrix is effectively reduced to $U \equiv U_{12} U_{13}$.

In the approximation we are discussing, $m^2$ is the only relevant
mass-scale difference and the vacuum
transition probabilities between different
neutrino flavours are:
\begin{eqnarray}
P(\nu_e\rightarrow\nu_\mu)&=&  4 \, s_{12}^2\, c_{12}^2\, c_{13}^4 
\,\sin^2\left(\frac{m^2\,L} {4\, E_\nu}\right) 
\cr
P(\nu_e\rightarrow\nu_\tau)&=& 4 \, s_{13}^2\, c_{13}^2\, c_{12}^2 \,
\sin^2\left( \frac{m^2\,L} {4\, E_\nu}   \right)
\cr
P(\nu_\mu\rightarrow\nu_\tau)&=& 4 \, s_{13}^2\, c_{13}^2\, s_{12}^2
\,\sin^2\left( \frac{m^2\,L} {4\, E_\nu} \right).
\label{todasprobs}
\end{eqnarray}
The probabilities
$P(\nu_e \rightarrow \nu_\mu)$ and $P(\nu_e \rightarrow \nu_\tau)$ are
quadratically suppressed for small $s_{12}$ and $s_{13}$, 
while $P(\nu_\mu \rightarrow \nu_\tau)$ is quartically suppressed.
The situation in  matter, however, is drastically different.

It is convenient to work in the ``kinetic'' basis wherein
$M^{vac}_K$ is diagonal. The effect of matter is fully encrypted
in a modification of the squared mass matrix:
\begin{equation}
M^{mat}_K \equiv M^{vac}_K + U^\dagger \left(\matrix{
A+B  & 0 & 0 \cr 
0  & B & 0 \cr
0 & 0 & B \cr}
\right) U,
\label{mmatter}
\end{equation}
where, as is well known,  $B$ arises from 
flavour-universal forward-scattering neutral current interactions while
$A$, given by Eq.~(\ref{Amat}), arises from the charged-current contribution specific to $\bar\nu_e\, e$
and $\nu_e\, e$ scattering.

To illustrate how matter effects lift the vacuum degeneracy between 
two mass eigenstates, we diagonalize $M^{mat}_K$  
to first order in $A/m^2$, temporarily assumed to be small (the exact
formulae, used in the numerical results, are presented in Appendix A). 
To order zero in this expansion there are two equal eigenvalues, so that 
we must follow the usual rules of degenerate perturbation theory.
Write $M^{mat}_K = M^{[0]} + M^{[1]}$ with:
\begin{eqnarray}
M^{[0]} & = & \left(\matrix{
0  & 0 & 0 \cr 
0 & m^2+ A\,  s_{12}^2 & A\,  c_{12} s_{12} s_{13} \cr
0 & A\,  c_{12} s_{12} s_{13} & m^2 + A\,  c_{12}^2 s_{13}^2 \cr}
\right)  \nonumber\\
M^{[1]} & = &\left(\matrix{
 A\,  c_{12}^2 c_{13}^2  & A\,  c_{12} c_{13} s_{12} &  A\,  c_{13} c_{12}^2 s_{13} \cr 
A\,  c_{12} c_{13} s_{12}  & 0 & 0 \cr
A\,  c_{13} c^2_{12} s_{13} & 0 & 0 \cr} 
\right)
\end{eqnarray}
where we have subtracted the common entry $B$, which plays no role
in neutrino mixing. We must diagonalize
$M^{[0]}$ exactly to lift
the degeneracy in $M^{vac}$, then the second term can 
be consistently treated in perturbation theory.

To order $A/m^2$,
the flavour and kinetic mass matrices, $M^{mat}_F$
and $M^{mat}_K$, are:  
\begin{eqnarray}
M^{mat}_F &= &U_{12} U_{13} U_{mat}\,M^{mat}_K\, 
                                U_{mat}^\dagger U_{13}^\dagger U_{12}^\dagger\; ; \\
M^{mat}_K &=&{\rm Diag}\;\left[\,A\,  c_{12}^2 c_{13}^2 ; \;
m^2 ; \; m^2 + A\,  ( s_{12}^2 + c_{12}^2 s_{13}^2)\,\right] 
\end{eqnarray}
where
\begin{equation}
U_{mat} \equiv 
\left(\matrix{
1  & 0 & 0 \cr 
0 & c^{mat}_{23} & s^{mat}_{23} \cr
0 & -s^{mat}_{23} &
 c^{mat}_{23} \cr}
\right),
\end{equation}
and the sine of the mixing angle in matter is:
\begin{equation}
s^{mat}_{23} = s_{12}/\sqrt{s_{12}^2 + c_{12}^2 s_{13}^2}\, 
\label{sineff}. 
\end{equation}

There are two important points to notice. First, instead of one 
mass difference as in vacuum, we have two:
\begin{eqnarray}
\Delta { m}^2_{12} & \sim & m^2\nonumber\\
\Delta { m}^2_{23} & = & A\, (s_{12}^2 + c_{12}^2 s_{13}^2)\; ,
\label{dm23m}
\end{eqnarray}
one of ${ O}(m^2)$, 
the other of ${ O}(A)$, the matter-induced mass. Secondly, the 
matrix $U_{mat}$ plays the same role 
as the mixing matrix $U_{23}$ in the generic mixing scenario
of three neutrinos. The 
effect of matter is simply to split the degenerate eigenvalues and induce the effective angle $s^{mat}_{23}$ of Eq.~(\ref{sineff}). The crucial
point is that this angle can be large even if $s_{12}$ and $s_{13}$  are small. To this order in $A/m^2$, the condition for maximal mixing is
$s_{13}={s_{12}}/{c_{12}}$;
 in a parametrization-independent language, this is equivalent to the requirement that the mixing between the second and third eigenstates with the electron flavour state be the same: 
$U(e2) = U(e3)$.

\subsection{Oscillation probabilities}

For the Earth's electron density appearing in $A$, 
Eq.~(\ref{Amat}),
it is a good approximation~\cite{density}
to consider a piecewise-constant density profile: 
a negligible density for neutrinos traversing the atmosphere,
$\rho_m = 5~ {\rm g/cm^3}$ for the mantle, and
$\rho_c = 11~ {\rm g/cm^3}$ for a core of 3500 km radius.
The matter-induced squared ``mass'' can then be expressed as 
 $A = 2 \sqrt{2} G_F E_\nu \langle n_e (c_\nu)
\rangle$, where the electron density is averaged over the 
neutrino trajectory, and $c_\nu$ is the cosine
of its zenith angle.
The oscillation probabilities depend  
on the two mass differences of Eq.~(\ref{dm23m}) 
and have the same form as the CP-conserving part of the 
general three-flavour vacuum-transition probabilities:
\begin{equation}
P_{\nu_\alpha\nu_\beta}(E_\nu, c_\nu)\,=\, 
-\,4\; \sum_{k>j}\,{\rm Re}[W_{\alpha\beta}^{jk}]\, 
\sin^2\left(
\frac{ \Delta m_{jk}^2 \, L(c_\nu)}{4\,E_\nu}
\right)\, ,
\label{reim}
\end{equation}
with $W_{\alpha\beta}^{jk}\equiv  \,
[U_{\alpha j} U_{\beta j}^* U_{\alpha k}^* U_{\beta k}]$, and 
$U \equiv U_{12}U_{13}U_{mat}$. The distance $L(c_\nu)$ is:
\begin{equation}
L(c_\nu) = R_\otimes 
( \sqrt{(1+l/R_\otimes)^2-s_\nu^2}-c_\nu),
\end{equation} 
where  $R_\otimes$ is the radius of the Earth and $l\sim 15$ km
is the typical height at which primary cosmic rays interact
in the atmosphere.

Consider the $\nu_\mu\leftrightarrow \nu_\tau$ entry of
 Eqs.~(\ref{reim}):
\begin{equation}
P_{\nu_\mu\nu_\tau}(E_\nu, c_\nu)\,=\, 
\sin^2(2\, \theta^{mat}_{23})\, 
\sin^2\left(
\frac{ \Delta m_{23}^2 \, L(c_\nu)}{4\,E_\nu}
\right)\, +\,{ O}(s^2_{12},s^2_{13}).
\label{numunutau}
\end{equation}

Even if  
$s_{12}$ and $s_{13}$ are small, $P_{\nu_\mu\nu_\tau}$ can be
maximal, since $\theta^{mat}_{23}\simeq \pi/4$
for $s_{12}\simeq s_{13}$.
In the limit  $s_{12},s_{13}\to 0$, 
$\Delta m_{23}^2\to 0$ and the oscillation probability vanishes:
there cannot be oscillations if all CKM angles are zero.  
  
Large $\nu_\mu\rightarrow \nu_\tau$ oscillations
take place for $s_{12}, s_{13}$ small, but bounded
from below by the condition 
$\Delta m^2_{23}~ R_\otimes/E_\nu \sim O(1)$.  
For this parameter range one should still check the size of 
$\nu_e \rightarrow \nu_\mu, \nu_\tau$ transitions,
which are not observed.
 This turns out not to be a problem for the atmospheric anomaly, 
 because the $\nu_\mu/\nu_e$ flux ratio  
is close to 2, and in the region of maximal mixing 
and small vacuum angles  $P(\nu_e \rightarrow \nu_\mu) 
\sim P(\nu_e \rightarrow \nu_\tau)$. Consequently,  
the number of disappearing $\nu_e$ 
can be compensated by the number of $\nu_\mu$s that 
oscillate into $\nu_e$s~\cite{trimaximal}. 
But a large 
$\nu_e$ disappearance probability can lead to a violation of 
the stringent bounds imposed by the Chooz experiment~\cite{chooz}
~\cite{declais}. 
We shall see that Chooz, but not SK, disfavours our
small angle scenario.

\subsection{Relation to the conventional scenario}

The degenerate-neutrino scenario
involves only one vacuum mass difference in the description of
terrestrial experiments. 
This is also the case in the scenarios
considered in most previous analyses of atmospheric data, even
with three families~\cite{three}, but with a single dominant mass
difference. 

As it turns out, the degenerate scenario is exactly equivalent to the conventional one of~\cite{three} with ${\bar \Delta m_{23}^2}=-\,m^2$ 
(in vacuum oscillations the sign of this 
difference would be unobservable).
To see this equivalence explicitly, it suffices to consider their vacuum CKM matrix,
which is written in the customary order
$\bar U \equiv {\bar U}_{23}({\bar s}_{23})\; {\bar U}_{13} ({\bar s}_{13})$,
and to obtain from it the matrix of Eq.~(\ref{ourCKM})
via the substitutions:
\begin{eqnarray}
{\bar s}^2_{23} &=&{ {s_{12}^2 c_{13}^2}\over
{s_{12}^2+s_{13}^2 c_{12}^2}}\, ,\cr
& & \cr
{\bar s}^2_{13} &=& c_{12}^2 c_{13}^2.
\end{eqnarray}

Note that small mixing angles in the degenerate
parametrization may correspond  to large mixings in the conventional one. 
In particular, a region of arbitrarily small mixing angles $s^2_{12}, s^2_{13}$ of our mass-degenerate scenario
is mapped to a domain around the values 
${\bar s}^2_{23} \sim 1/2$ and ${\bar s}^2_{13} \sim 1$ of the
conventional parametrization.
The most ``natural'' parametrizations are the ones in
which the rotation matrices  $U_{ij}$ act on the
mass eigenstates in  order of decreasing degeneracy.
The conventional parametrization, used in the Particle Data Book~\cite{PDG},
 is natural for the quark sector with its hierarchical
mass splitting, but not necessarily for the lepton sector. The parametrization we use, Eq.~(\ref{ourCKM}), is natural for the partially-degenerate mass pattern that we are considering~\cite{tanos}. 

As we saw in the previous subsection, the presence of degenerate eigenstates in vacuum can lead to large transition probabilities in matter.  The enhacement of transition probabilities in matter in the context of three-family mixing 
with two degenerate neutrinos has been discussed before in \cite{degshift} and, in the context of the three-maximal mixing model, in \cite{trimaximal}. 
The parametrization we use here  
clarifies the origin and generality of the effect.

\section{Zenith angle and energy distribution of the SK events}

The data of the SK collaboration, as well as their Monte Carlo
expectations for the case in which there are no neutrino oscillations,
are binned in the azimuthal angle of the observed
electrons and muons, and in their energy (in the case of muons
the level of containment within the detector also distinguishes
different data samples). To reproduce these results one must convolute
neutrino fluxes and survival probabilities with charged-current
differential cross sections and implement various efficiencies
and cuts. This being an elaborate procedure, in Appendix D we 
check our results by reproducing the no-oscillation Monte Carlo 
results of SK, as well as the neutrino ``parent energy'' spectrum:
the azimuthally averaged neutrino flux weighted with the integrated
neutrino cross section and with the efficiencies of the various
data samples. 

In the rest of this section we review how the data are binned,
we specify our procedure, and we analyze the fits to the
conventional oscillation scenario, as well as to our mass-degenerate
alternative.

\subsection{The data samples}

The SK collaboration chooses to bin the observed charged-lepton energies in a few samples. 
The electron candidates are subdivided
into sub-GeV (sgev) and multi-GeV (mgev). The muon candidates
are distinguished as sgev and mgev, partially and fully contained (PC, FC),
and through-going (thru). To set apart these categories, we introduce
selection functions ${\rm Th}_{s,l }(E_l,c_l)$, with $s=$ sgev, mgev and $l=e,\mu$,
that depend on the energy, $E_l$, and on the cosine, $c_l$, of the
azimuthal angle of the outgoing lepton ($c_l=1$ is vertically down-going).
In computing the number of events, these selection functions will weight
the product of neutrino flux and cross-section.
For the sgev events,  
\begin{eqnarray}
 {\rm Th}_{{\rm sgev},l }(E_l) \equiv  \Theta[ E_{{\rm th},l } - E_l]\; 
\Theta[E_l-E_{\rm min,l }],
\end{eqnarray}
with $E_{{\rm th}, e(\mu)}=1.33 \, (1.4)$ GeV, $ E_{{\rm min}, e(\mu)} = 100 \, (200)$ MeV for $e$($\mu$) respectively. For the mgev electron events, ${\rm Th}_{{\rm mgev},e}(E_l)  \equiv  \Theta[ E_l - E_{{\rm th},e}]$.

For the mgev muons, we must distinguish between partially 
and fully contained events:
\begin{eqnarray}
{\rm Th}_{{\rm mgev-PC},\mu}(E_\mu,c_\mu)  \equiv  \Theta[E_\mu - E_{{\rm th},\mu}]\; {\rm PC}(E_\mu,c_\mu)\, ,\nonumber\\
{\rm Th}_{{\rm mgev-FC},\mu}(E_\mu,c_\mu)  \equiv  \Theta[E_\mu - E_{{\rm th},\mu}]\; {\rm FC}(E_\mu,c_\mu)\,,
\end{eqnarray}
where the functions ${\rm FC}(E_\mu,c_\mu)$, ${\rm PC}(E_\mu,c_\mu)$ measure the fraction of the total 
fiducial volume in which a neutrino interaction could produce a 
$\mu$ with energy $E_\mu$ and zenith direction $c_\mu$ that either stops 
before exiting the detector (FC) or escapes (PC). We have explicitly
constructed ${\rm FC}(E_\mu,c_\mu)$ and ${\rm PC}(E_\mu,c_\mu)$ using the shape and size of the detector and
 the $\mu$ range in water, $R_w(E_\mu)$, as a function of energy, and 
describe this in
 Appendix C.

We have also devised a through-going muon selection function 
${\rm Th}_{\rm thru,\mu}(E_\mu,c_\mu)$. 
The effective target mass of the rock surrounding SK depends on 
energy via the muon range in water
and in rock, $R_r(E_\mu)$. 
The observed muon energy is required to be 
greater than $E_{\rm min}'=1.6$ GeV, implying that its trajectory must be
longer than 7 m. The function ${\rm Th}_{\rm thru,\mu}$ must account
for the detector's effective area for such tracks,  
${\rm A}(E_\mu, c_\mu)$, which depends, via the muon range,
on the muon energy as
it enters the detector, and on its zenith angle. 
Furthermore, the selection function for through-going muons
must take into account that their flux, as given by SK, is defined as  
the number of events divided by the
effective area for a muon of energy $E_{min}'$~\cite{skup}. 
All in all:
\begin{equation}
{\rm Th}_{\rm thru,\mu}(E_\mu,c_\mu)\equiv
[R_r(E_\mu) -R_r(E_{min}')] \; \frac{{\rm A}(E_\mu, c_\mu)}{{\rm A}(E_{min}', c_\mu)}.
\label{thru}
\end{equation}

This effective area is given in Appendix C.

\subsection{Number of events}

Let $d\Phi_{\nu}(E_\nu, c_\nu)/dE_\nu\,dc_\nu$ 
and $d\bar\Phi_{\nu}(E_\nu, c_\nu)/dE_\nu\,dc_\nu$ be the atmospheric 
neutrino fluxes of $\nu=\nu_e,\nu_\mu$ and their antiparticles,
with $E_\nu$ the neutrino energy and $c_\nu$ its zenith angle (we use
the Bartol code \cite{bartol} of atmospheric neutrino fluxes at the 
Kamiokande site).
Let $d\sigma(E_\nu, E_l, c_\beta)/dE_l\,dc_\beta$ and 
${d \bar \sigma}(E_\nu, E_l, c_\beta)/dE_l\,dc_\beta$ be the neutrino and antineutrino charged-current cross sections, which depend on $E_\nu$, 
on the outgoing lepton energy $E_l$, and on the cosine of the scattering angle between the 
two particles, $c_\beta$. In Appendix B, we discuss in detail the 
cross sections used in our analysis. The zenith angle of 
the outgoing lepton $c_l$,  which is the measured quantity,  
is a function of  $c_\nu$, $c_\beta$, and of the azimuthal angle, $\phi$, of the outcoming lepton in the target rest frame. 

Let  $N^{0}_{s,l}(c), N^{{\rm osc}}_{s,l}(c)$ be the expected number of
charged-current events in the sample $s$ ($s=$ sgev, mgev-pc, mgev-fc, thru) for $l=e, \mu$ and in the bin in zenith cosine with central value $c$, for the no-oscillation (0) and oscillation (osc) hypotheses. For the sgev and 
mgev samples, the theoretical prediction is given by: 
\begin{eqnarray}
 N^{\rm osc}_{s,l}(c)= K_{s,l}  \int  dE_\nu\, dc_\nu\, dE_l\, dc_\beta\, d\phi  \; {\rm Th_s}(E_l,c_l)\; (\Theta[c_l-c+\delta] -\Theta[c_l-c-\delta]) 
\;\;\;\;\;\;\nonumber\\
\!\!\!
\sum_{\nu'=\nu_e,\nu_\mu} \left[ \sigma(E_\nu, E_l, c_\beta) \frac{d\Phi_{\nu'}}{dE_\nu dc_\nu}  \, P_{\nu'\nu}(E_\nu, c_\nu)  +
 {\bar \sigma}(E_\nu, E_l, c_\beta) \frac{d{\bar\Phi}_{\nu'}}{dE_\nu dc_\nu}\, {\bar P}_{\nu'\nu}(E_\nu, c_\nu)\right] \nonumber\\
\label{chori2}
\end{eqnarray}
for the oscillation hypothesis, with ${P}_{\nu'\nu}(E_\nu, c_\nu)$
 and ${\bar P}_{\nu'\nu}(E_\nu, c_\nu)$ the oscillation probabilities from flavour
$\nu'$ to flavour $\nu$ for neutrinos and antineutrinos respectively. To obtain $N^{0}$, take 
${P}_{\nu'\nu}={\bar P}_{\nu'\nu}=\delta_{\nu'\nu}$.
The $\Theta$ functions in Eq.~(\ref{chori2}) express the constraint
that $c_l$ be in the bin with
central value $c$ and width $2 \delta = 0.4$,  the binning used in SK for the sgev and mgev samples. Finally,
$K_{s,l}$ are normalization constants, which 
ensure that the total number of events in each sample is the same 
as in the SK Monte Carlo data  for the non-oscillation hypothesis. By choosing these factors by hand,
we skirt the question of efficiencies
for electron or muon detection and for single- or multiple-ring events:
all we need to assume is that they are roughly constant
within a given data sample, which we believe to be the case. 
We neglect the cross-talk between different samples. 

For the flux of through-going muons, we have:
\begin{eqnarray}
\Phi^{\rm osc}_{\rm thru}(c)  =  K_{\rm thru} 
\int d\phi\, dc_\beta\, dE_\mu dc_\nu\, dE_\nu \; 
{\rm Th}_{\rm thru,\mu}(E_\mu,c_\mu) \;
(\Theta[c_\mu-c+\delta] -\Theta[c_\mu-c-\delta])\nonumber\\
\sum_{\nu'=\nu_e,\nu_\mu} \left[\sigma(E_\nu, E_\mu, c_\beta) 
\frac{d\Phi_{\nu'}}{dE_\nu dc_\nu} P_{\nu' \nu_\mu}(E_\nu, c_\nu)  + {\bar\sigma}(E_\nu, E_\mu, c_\beta) \;\frac{d{\bar\Phi}_{\nu'}}{dE_\nu dc_\nu} 
{\bar P}_{\nu'\nu_\mu}(E_\nu, c_\nu) \right]\nonumber\\
 \;\;\;\;\;
\end{eqnarray}
for the case with oscillations.
The width of the zenith angle bins in this sample is $2\delta = 0.1$.

\subsection{Results of the analysis of the SK data}

We have performed a $\chi^2$ analysis of the oscillation hypothesis 
in our mass-degenerate scenario, for both signs of $m^2$, using the 
full 52 kiloton-year
 data sample gathered by the Super-Kamiokande collaboration 
in 848 days of exposure \cite{kate}.  
The case of negative $m^2$ 
is exactly equivalent to the conventional scenario considered in \cite{fogli}, 
as shown in Section 2. The measured quantities are the 30 zenith angle bins
measured by SK in the five types of data samples. 
The choice of an error correlation matrix is  
non-trivial, as there are large theoretical uncertainties in 
the input neutrino flux, which induce large correlations between the
errors in the different measured quantities. We have constructed the
error correlation matrix in the same way as the authors of~\cite{fogli}, to
whom we refer for details. To gauge the incidence
of these ``systematic'' errors on the results, we have also  
performed the analysis with only statistical uncertainties. 

\begin{figure}[htb]
  \centering
\mbox{\epsfig{file=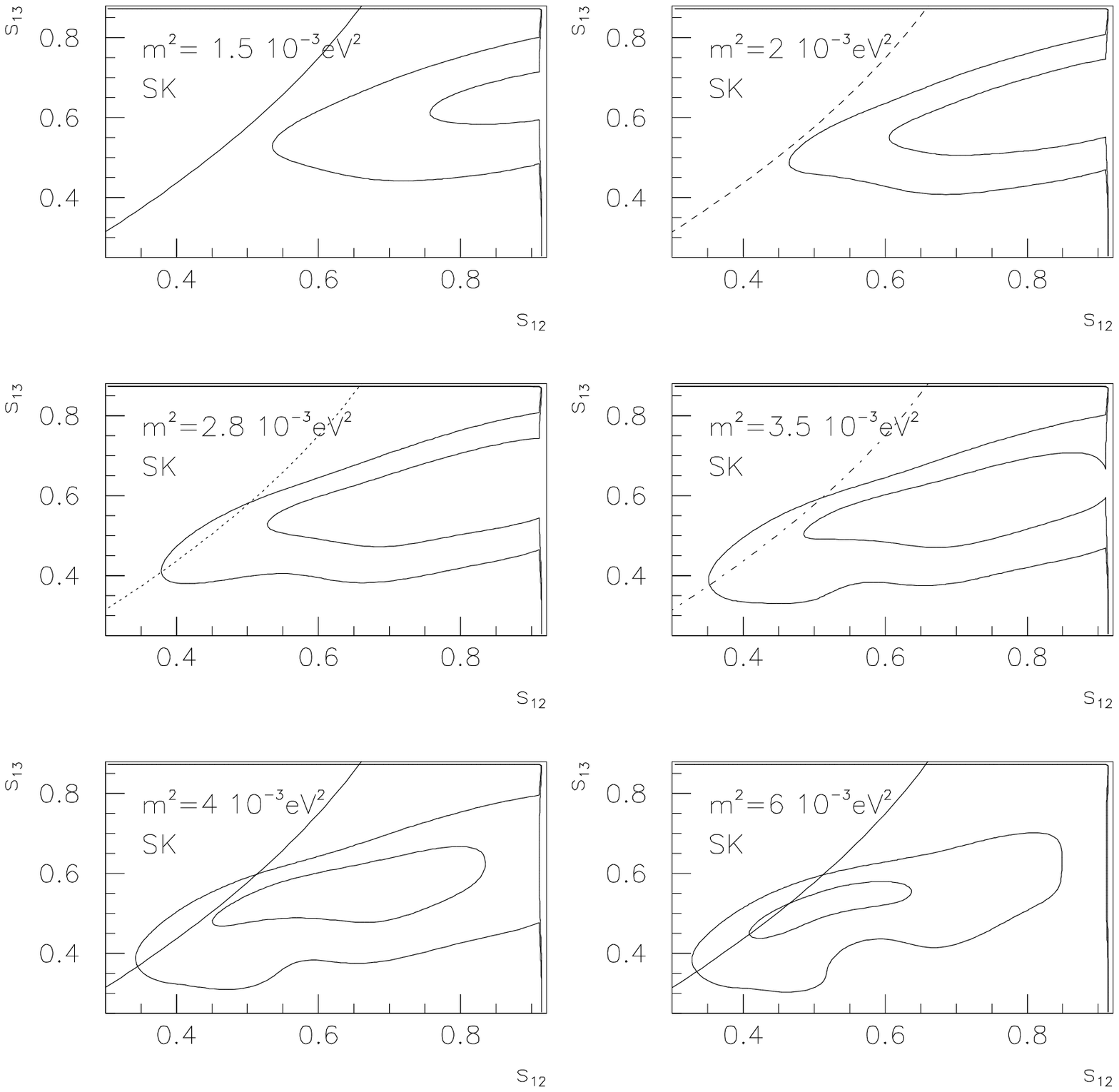,width=4.in,height=3.7in}}
  \caption{$68.5, 99\%$ CL intervals allowed by SK data alone,
for different values of $m^2$.} 
\label{sk1}
\end{figure}

\begin{figure}[htb]
  \centering
\mbox{\epsfig{file=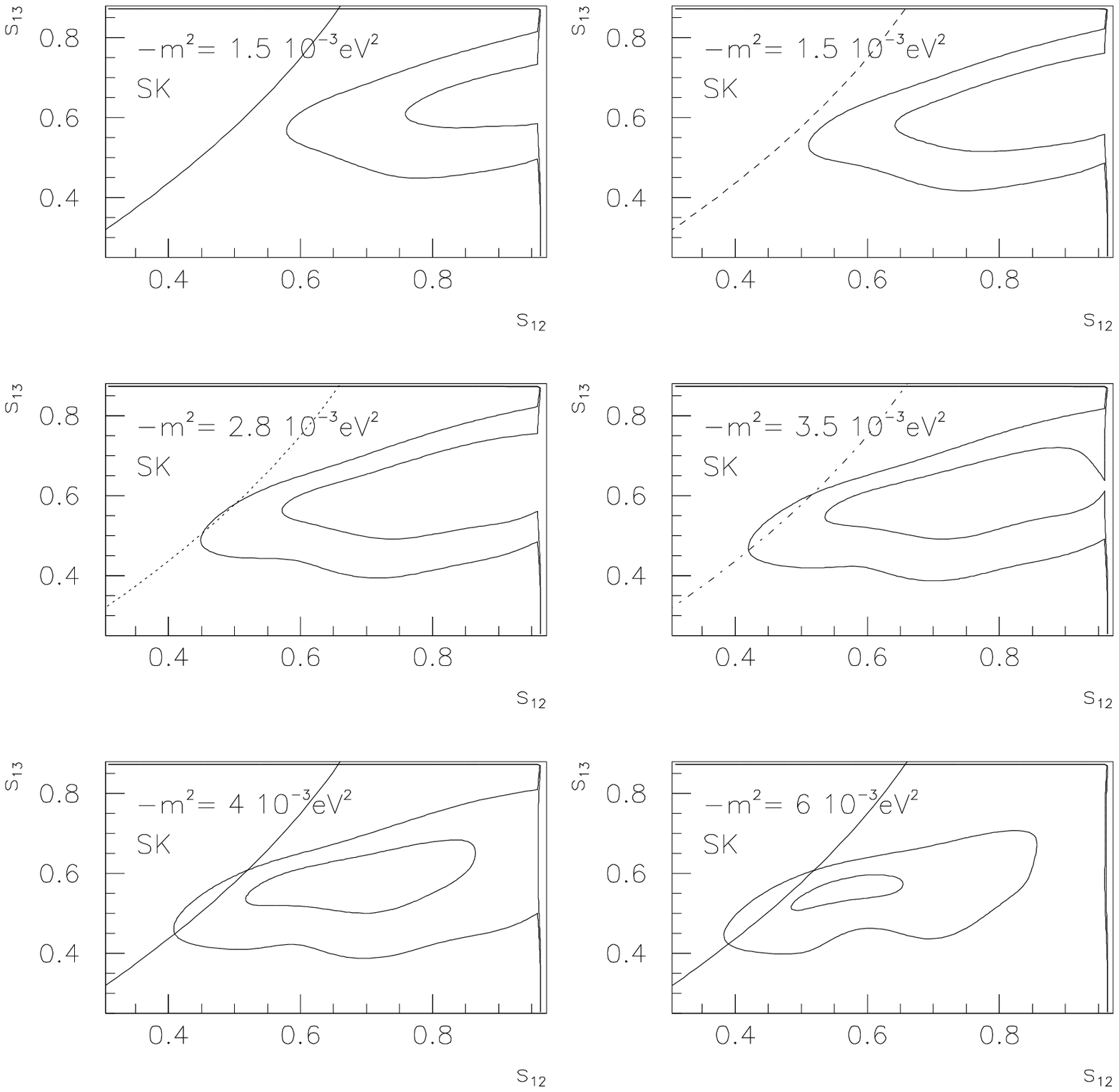,width=4.in,height=3.7in}}
  \caption{The same as Fig.~\ref{sk1},  but for negative $m^2$.}
\label{sk2}
\end{figure}

In Fig.~\ref{sk1} we show,  in the 
plane $s_{12}$--$s_{13}$ and for several positive values of  $m^2$,
the contour lines delimiting 
the allowed regions at 68.5 and 99\% confidence.
In Fig.~\ref{sk2} the same information
is displayed for negative $m^2$. The region of maximal mixing
in the conventional parametrization corresponds to values
$s_{12}\sim 1$ and $s_{13} \sim 1/\sqrt{2}$ of our parametrization.
This region is favoured for
the smaller allowed values of $m^2$: the top two rows of
Fig.~\ref{sk1}.  At the larger values of 
$m^2$, however, the contours extend largely to a region with
significantly smaller
vacuum angles,  the oscillation probabilities 
being enhanced by matter effects. 
We draw for comparison the line corresponding to maximal mixing 
in the perturbative approximation of Eq.~(\ref{sineff}),  valid
for the larger $m^2$ values. The allowed regions
 at small angles are close to this line, as expected. At
values of $m^2$ smaller than those shown in the figure,
the allowed regions shrink around the conventional 
maximal-mixing solution. 

The minimum $\chi^2$ is obtained for $|m^2| \sim 3.5\times 10^{-3}$ eV$^2$,
independently of whether the errors are taken to be purely statistical, or
estimates of flux uncertainties are also included.
This result is in good agreement with
that found by the SK collaboration in a
two-family mixing context. We do not find such an agreement
on the optimized mixing angles.
The best-fit angles in our parametrization are $s_{12}^2 \sim 0.42~(0.45)$ and $s_{13}^2 \sim 0.31~(0.33)$ for positive (negative) $m^2$, which in the conventional parametrization 
correspond 
to ${\bar s}_{23}^2 = 0.48~(0.48)$ and ${\bar s}_{13}^2 = 0.4~(0.37)$,  not so
far from the so-called tri-maximal mixing model ($s_{12}^2 =1/2, s_{13}^2=1/3$) \cite{trimaximal}. The Chooz data, however, disfavour 
these relatively large-$m^2$ and small-angle solutions.

\section{Chooz Constraints}

The reactor experiment Chooz provides tight upper limits
on the ${\bar \nu}_e$  disappearance probability in a domain with 
$\Delta m^2 \geq 10^{-3}~{\rm eV}^2$~\cite{chooz}\cite{declais}.
This entails very strong strictures on  
$P_{{\bar \nu}_e {\bar \nu}_\mu}$ and $P_{{\bar \nu}_e \bar{\nu}_\tau}$  
at atmospheric distances. 

We have constrained our analysis of atmospheric data to
comply with the Chooz results on the ratio $R$ of 
observed $e^+$ events to the number expected in the absence of
oscillations:
\begin{eqnarray}
R = \frac{\int d E_\nu \; \Phi(E_\nu) \; \sigma(E_\nu) \; 
P_{{\bar \nu}_e {\bar \nu}_e}(E_\nu)}{\int d E_\nu \; 
\Phi(E_\nu) \; \sigma(E_\nu)}\, , 
\label{esa}
\end{eqnarray}
where $\Phi(E_\nu)$ is the spectrum of neutrinos, obtained by combining,
in the appropriate proportions~\cite{declais}, the decay spectra 
of the different isotopes in the Chooz reactors~\cite{fluxes}. 
In writing Eq.~(\ref{esa}) we have approximated the efficiency
as a constant, for lack of better information.
The cross section, $\sigma(E_\nu)$, including the threshold effects, has been obtained from \cite{vogel}. For the transition probabilities 
we can use 
Eq.~(\ref{todasprobs}), since matter effects are completely negligible. 

\begin{figure}[htb]
  \centering
\mbox{\epsfig{file=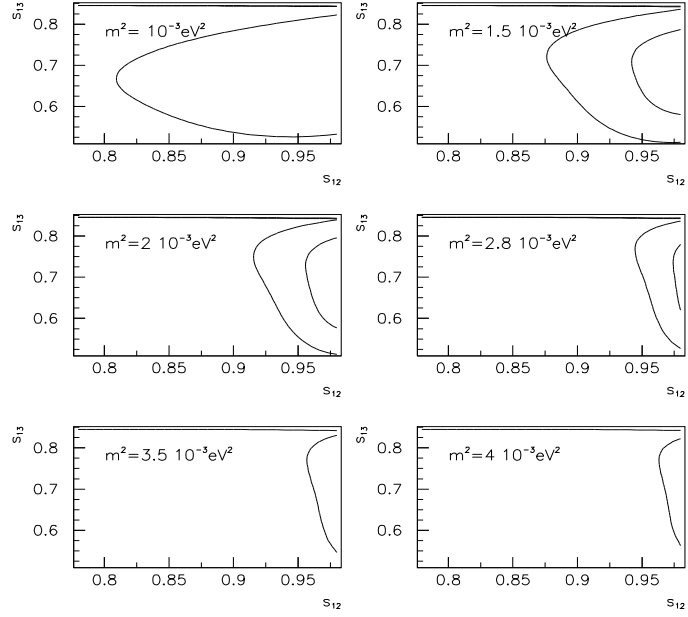,width=4.in,height=3.7in}}
  \caption{$68.5, 99\%$ CL intervals allowed by SK 
and Chooz data for different values of $m^2$. }
\label{skchooz1}
\end{figure}

\begin{figure}[htb]
  \centering
\mbox{\epsfig{file=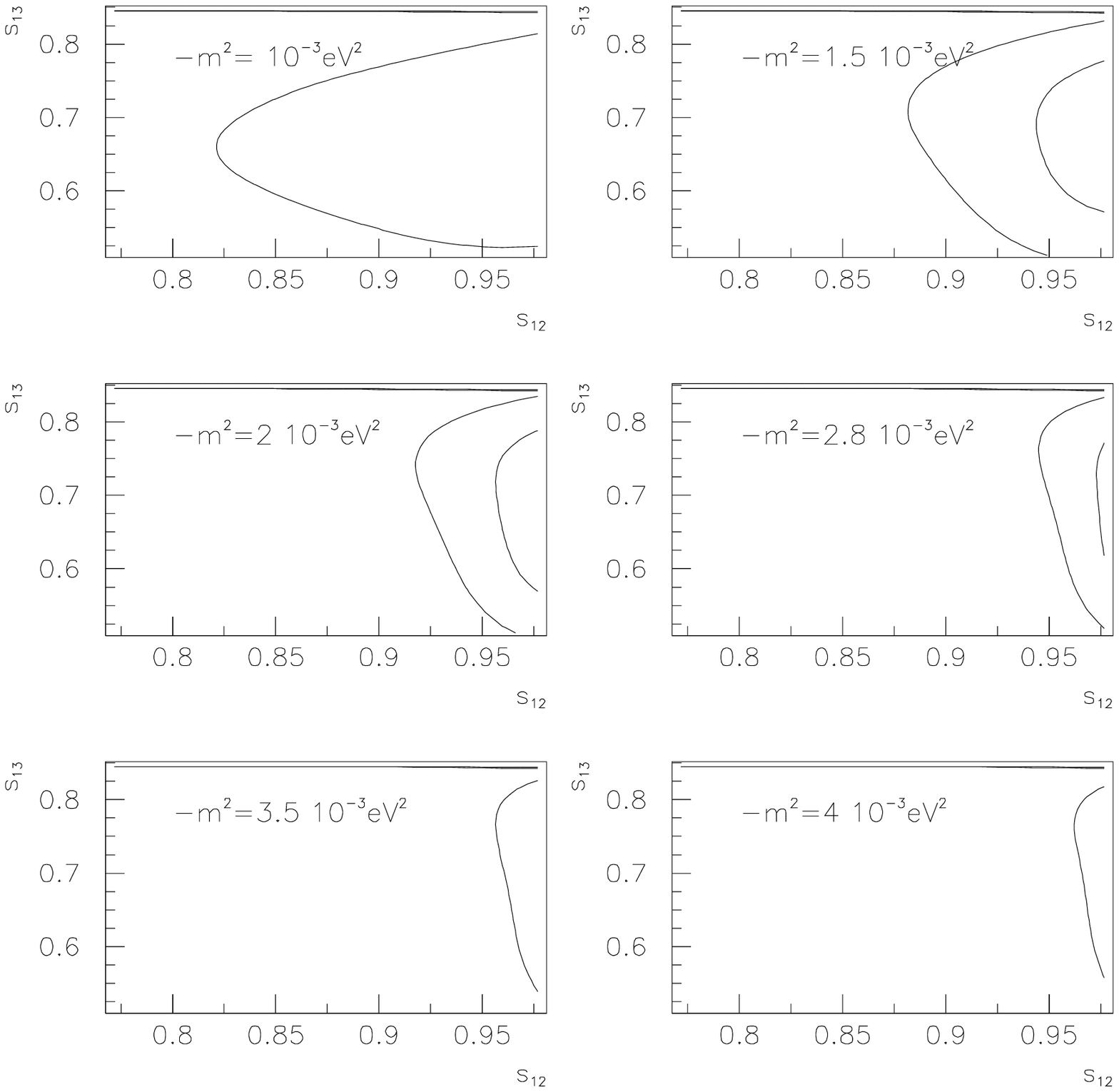,width=4.in,height=3.7in}}
  \caption{The same as Fig.~\ref{skchooz1} but for negative $m^2$.}
\label{skchooz2}
\end{figure}

The results of this combined SK--Chooz analysis are shown in 
Fig.~\ref{skchooz1} for positive $m^2$. The results for negative $m^2$ shown
in Fig.~\ref{skchooz2} are very similar. Clearly the Chooz data favour the conventional 
maximal-mixing solution as the only acceptable one.

\begin{figure}[htb]
  \centering
\mbox{\epsfig{file=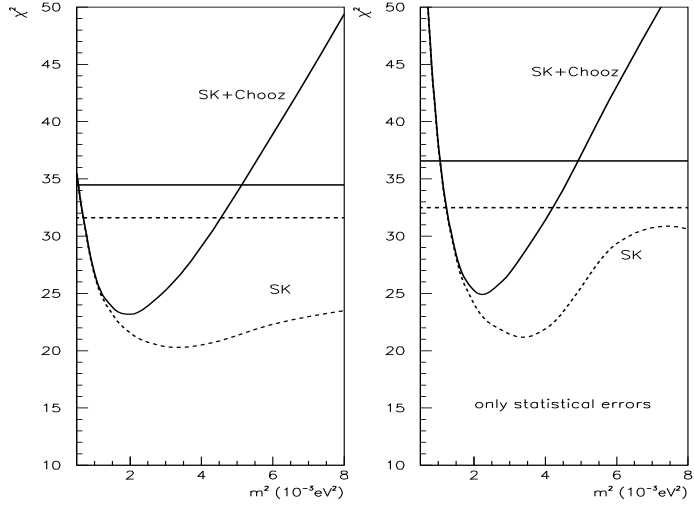,width=4.in,height=3.in}}
  \caption{Minimum $\chi^2$ as a function of $m^2$. The solid line 
corresponds to SK+Chooz data, while the dashed line includes only SK data. The
curves on the left plot include theoretical uncertainties as in \cite{fogli}, while the one on the right includes only statistical errors. The horizontal lines correspond to $99\%$ CL intervals for three
degrees of freedom.}
\label{masschi1}
\end{figure}

\begin{figure}[htb]
  \centering
\mbox{\epsfig{file=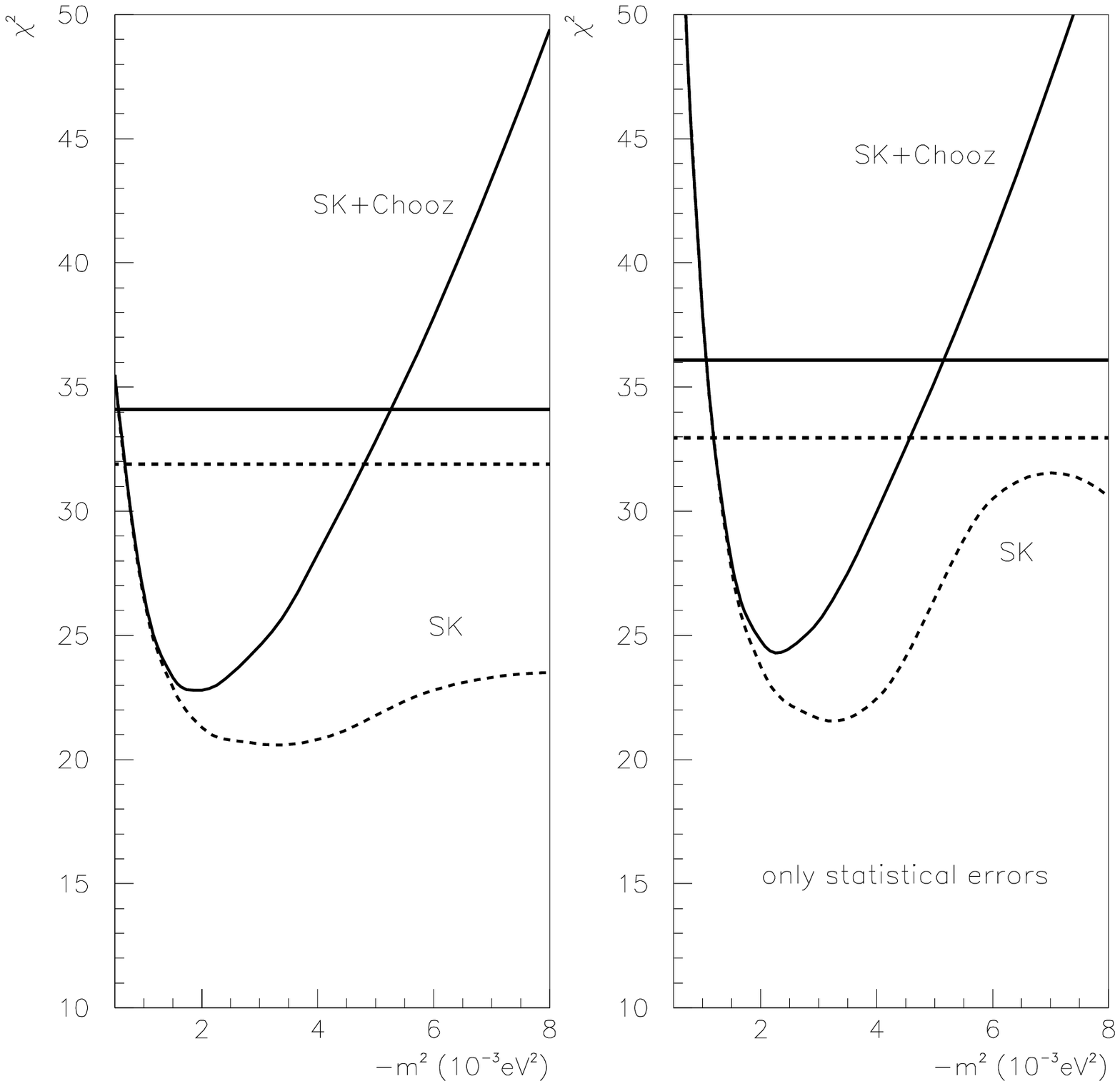,width=4.in,height=3.in}}
  \caption{The same as Fig.~\ref{masschi1} but for negative $m^2$.}
\label{masschi2}
\end{figure}

In Fig.~\ref{masschi1}, we show the minimum $\chi^2$ as a function of 
mass, for positive $m^2$. On the left we include
theoretical flux errors as in~\cite{fogli}, while on the right 
only statistical uncertainties are taken into account. Reassuringly,
the theoretical errors  have a small incidence on the results.
In Fig.~\ref{masschi2}, we show the results for negative $m^2$. 
For both signs of the mass difference, the minimum of the $\chi^2$ 
occurs at $|m^2| = 2$--$2.5\times 10^{-3}$ eV$^2$, which is slightly smaller 
than the value obtained in the combined analysis of~\cite{fogli} 
($|m^2| \sim 2.8\times 10^{-3}$ eV$^2$). 
Concerning the mixing amplitudes, we find as the best fit for the
important~\cite{dgh}~\cite{tanos}
 angle ${\bar \theta}_{13}$ --in the conventional
parametrization-- at ${\bar \theta}_{13} = 6^{\rm o}$, to be
 compared to $8^{\rm o}$ found in~\cite{fogli}.
However, the $\chi^2$ curve is flat enough for
${\bar \theta}_{13} = 0$ to be perfectly compatible with the data.

\begin{figure}[htb]
  \centering
\mbox{\epsfig{file=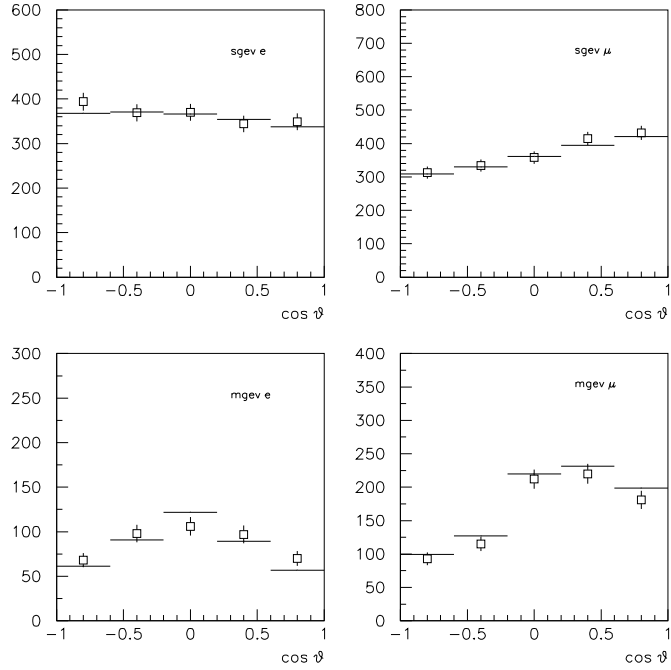,width=4.in,height=4.in}}
  \caption{Zenith angle distributions of $e$ and $\mu$ SuperKamiokande samples (squares)
compared with the best-fit oscillation hypothesis: 
$m^2 = 2\times 10^{-3}$ eV$^2$ and close to maximal mixing. The errors shown are only statistical.}
\label{zen_bf}
\end{figure}

\begin{figure}[htb]
  \centering
\mbox{\epsfig{file=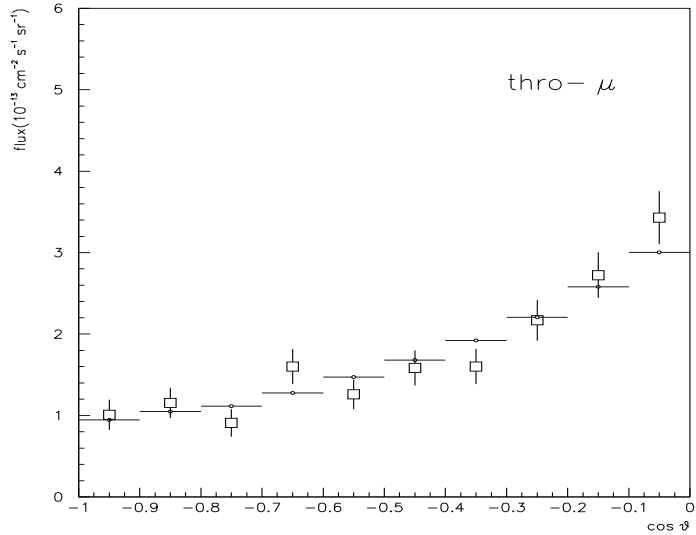,width=4.in,height=3.2in}}
  \caption{Zenith angle distribution of through-going muons (squares)
compared with the best fit oscillation hypothesis: $m^2 = 2\times 10^{-3}$ eV$^2$ and close to maximal mixing.}
\label{zen_bf_up}
\end{figure}

In Figs.~\ref{zen_bf} and \ref{zen_bf_up} we show the impressive agreement between  the SK zenith angle distributions
and our best-fit oscillation hypothesis, obtained including the Chooz constraint.

Incidentally, for the trimaximal mixing model we get $\chi^2=42\, (44)$ (for 30 degrees of freedom: 31 data minus one free parameter $m^2$) for positive (negative) $m^2$ at 
$|m^2| = 10^{-3}$ eV$^2$, a mass value for which the Chooz constraint is inoperative. The $\chi^2$ rises rapidly for larger 
$|m^2|$. The probability that this model is correct is below $10\%$.

\clearpage

\section{Conclusions}

The neutrino squared mass difference used to explain the 
atmospheric neutrino
anomaly in terms of oscillations is of the same order of magnitude
as that induced by Earth matter effects, for a typical atmospheric
neutrino energy. Triggered by this coincidence,
we set out to study --in a scheme with three neutrinos
and in minute detail-- whether or not the large depletion
of muon neutrinos observed by SK
could be due, not to {\it ab initio} large mixing angles, 
but to matter-enhanced smaller-angle mixings. 
Our suspicion turned out to be correct: the SK data can be very
satisfactorily explained with mixing angles that
are far from maximal. But the constraints from Chooz on the
survival of electron antineutrinos disfavour our non-maximal solution.
The conclusion that vacuum $\nu_\mu \leftrightarrow \nu_\tau$ neutrino 
transition probabilities are nearly maximal may be surprising,
but it is here to stay.

{\bf Acknowledgements}

\noindent
  We have had useful conversations with A. Donini, J. Ellis, 
F. Feruglio, G.L. Fogli, J. G\'omez-Cadenas, M.C. Gonzalez-Garc\'{\i}a, Y. Hayato, E. Lisi, M. Lusignoli, S. Rigolin and O. Yasuda. We are indebted to the Bartol and Honda groups for providing us with their atmospheric neutrino 
fluxes.  We warmly thank K.   
Scholberg  for many illuminating discussions on the SK data
and its interpretation. After 
the completion of this work, E. Lisi kindly pointed out to us the references \cite{degshift}, where matter effects in three--family mixing were previously 
discussed, and O. Peres kindly pointed out a missprint in eq. (8). The work of M.B.G. was 
also partially supported by CICYT project AEN/97/1678.

\vskip 1.5 cm 

{\Large{\bf Appendix A: Oscillation parameters}} 

The exact 
diagonalization of the mixing matrix in Eq.~(\ref{mmatter})
results in the effective eigen-mass differences:
\begin{eqnarray}
\Delta { m}^2_{12} & = &  \frac{1}{2} \left( m^2 - A + \sqrt{(m^2 + A )^2 - 
4~ A~ m^2~ c_{13}^2~ c_{12}^2}\right)\,,\nonumber\\
\Delta { m}^2_{23} & = &  \frac{1}{2} \left(-m^2 + A + \sqrt{(m^2 + A )^2 - 
4 A~ m^2~ c_{13}^2~ c_{12}^2}\right)\,,
\label{dm23exact}
\end{eqnarray}
and in  a mixing matrix:
\begin{eqnarray}
U_{mat} = \left(\matrix{
-(c_{12} c_{13} - a_+)/b_+  & 0 & ( c_{12} c_{13} - a_-)/b_- \cr 
-s_{12}/b_+ & s_{13} c_{12}/\sqrt{s_{12}^2+s_{13}^2 c_{12}^2} & s_{12}/b_-  \cr
-s_{13} c_{12}/b_+ & -s_{12}/\sqrt{s_{12}^2+s_{13}^2 c_{12}^2} & 
 s_{13} c_{12}/b_- \cr}
\right), 
\label{umatexact}
\end{eqnarray}
where 
\begin{eqnarray}
a_{\pm} & \equiv & \left ( A + m^2 \pm \sqrt{(m^2 + A )^2 - 
4~ A~ m^2~ c^2_{13}~ c^2_{12}}\right ) /(2~ A~ c_{12}~ c_{13}) ,\nonumber\\
b_{\pm} 
& \equiv & \sqrt{1 + a_{\pm}^2 - 2~ a_{\pm}~ c_{13}~ c_{12}}\; . 
\end{eqnarray} 

\vskip 0.5 cm

{\Large{\bf Appendix B: Neutrino cross sections} }

The proper treatment of neutrino cross sections
in water and rock, at energies ranging from 100 MeV to hundreds 
of GeV, is an arduous art. We do not attempt  an
elegant and complete analysis. Instead, we use a treatment
that --notwithsanding its oversimplifications-- is capable of
reproducing to an adequate level the observed total cross sections
and scattering angle distributions, which are the 
ingredients needed for the data analysis.

We simplify the neutrino cross sections in water 
by considering only oxygen as a target,
an isoscalar nucleus for which we ignore shadowing,
but not Fermi-motion effects, which we treat as 
in~\cite{lls}. We also neglect the
muon mass. As the SK experimenters do, we build a cross section 
out of three dominant contributions:
quasi-elastic ($\nu_l {\cal N}\to l {\cal N'}$),
resonant one-pion production, and ``deep'' inelastic.
In so doing we ignore the small contribution of the
``diffractive'' domain of relatively high energy, low $Q^2$. 

For the quasi-elastic cross section we use the standard
expression reviewed in \cite{lls}, with $M_A=1.0$ GeV
for the mass describing the axial form factor. For 
one-pion production we use Eq.~(22) of~\cite{blls} for
the excitation of the $N^*(1236)$ resonance of spin
and isospin 3/2. We assume that these contributions 
saturate the cross section for an invariant mass of
the final hadrons $W\leq 1.4$ GeV. Above that value
we use a deep-inelastic cross section with an 
exact Callan-Treiman constraint $F_2=2xF_1$. For the
structure functions $F_2$ and $F_3$ we use the
compilation of~\cite{grv94}. As an excuse to extend
this deep inelastic cross section to values of $Q^2$
as low as 0.4~GeV$^2$, we use $\xi$-scaling. This
is known to deal correctly with the higher-twist
target-mass corrections~\cite{Otto} and to interpolate
the resonant contributions in the sense of local
duality~\cite{dual}. Alas, the structure functions 
 are not extracted from the data using $\xi$-scaling. But the authors of~\cite{stirling}
have shown that, at least in the case of electroproduction,
a blind a posteriori use of $\xi$-scaling improves 
the fit to the lower $Q^2$ data (their prescription consists
in using the Bjorken-scaling cross-section expressions~\cite{PDG}
and the structure functions extracted at higher $Q^2$, with
the simple and not fully consistent a posteriori substitution $x\to\xi$
in the argument of the structure functions).
 
\begin{figure}[htb]
  \centering
\mbox{\epsfig{file=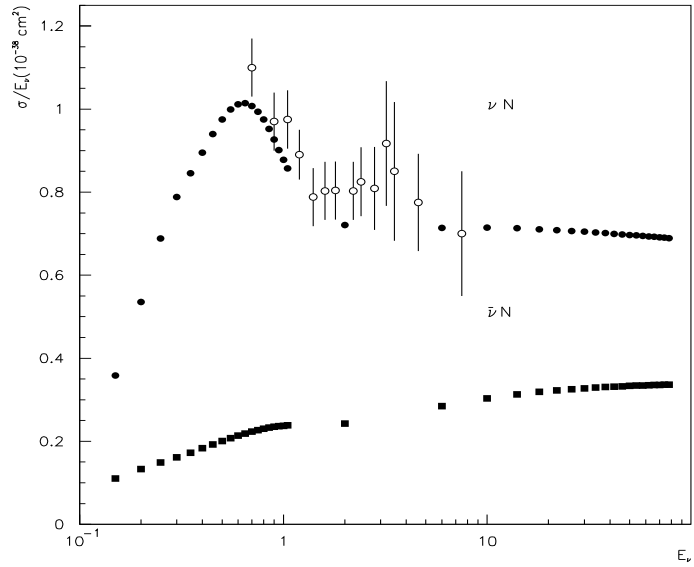,width=4.in,height=3.4in}}
  \caption{Total cross section over neutrino energy for $\nu$ and $\bar \nu$ 
charged-current scattering on an isoscalar target.}
\label{sigma}
\end{figure}

In Fig.~\ref{sigma}, we plot $\sigma_{_{\rm TOT}}(E_\nu)$ as a function of 
$E_\nu$ for neutrinos and antineutrinos. 
The curves are a bit below the available data at low energies, no doubt reflecting
the absence of a calculated diffractive contribution. The
SK collaborators, as well as many other authors,
extend the structure functions down to $Q^2=0$, thereby obtaining a slightly 
better ``fit'' to the $\sigma_{_{\rm TOT}}$ data. Rather than indulging 
in this inconsistent use of the deep inelastic structure functions, 
we have checked that our results on neutrino-mixing parameters
are insensitive to this kind of variations of the input. 

\begin{figure}[htb]
  \centering
\mbox{\epsfig{file=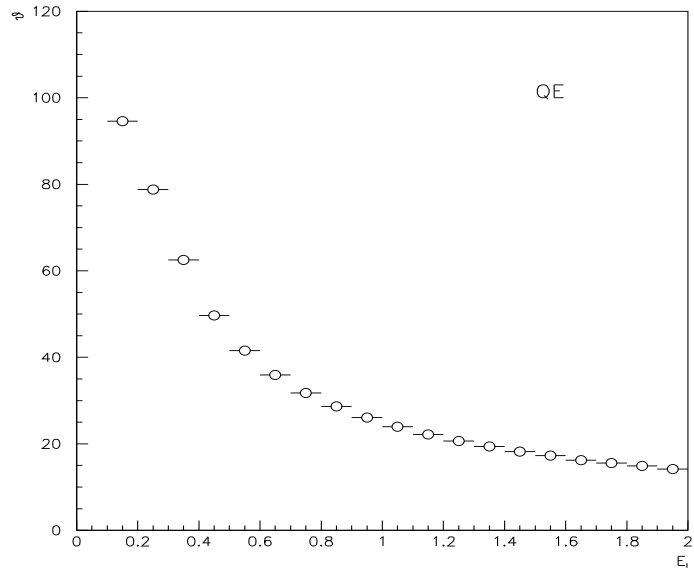,width=4.in,height=3.4in}}
  \caption{Average scattering angle between the parent neutrino and the lepton for quasi-elastic events.}
\label{cos}
\end{figure}

An important quantity in the zenith angle analysis is the average 
scattering angle between the lepton and the parent neutrino. In Fig.~\ref{cos}, we show the average angle for quasi-elastic events as a function of the 
lepton energy. This curve is in perfect agreement with that obtained
by the SK collaboration~\cite{okumura,hayato}.

\vskip 0.5 cm

{\Large{\bf Appendix C: Geometrical acceptances}}

The SuperKamiokande detector is a cylinder of height $H=36.2$ m
and radius $R=16.9$ m, of fiducial volume 
$V=\pi\,(R-2~{\rm m})(H-4~{\rm m})$. Let a point within
the detector be labelled by $y$, the height from the
bottom plane; $z$, the distance from the axis; and $x$,
a third cartesian coordinate.
Let $d(x,y,z)$ be the minimum distance from a point 
in the detector to its wall, let $s_\mu,c_\mu,t_\mu$ be the
sine, cosine and tangent of the muon's azimuthal 
angle
and let $E_{min}=0.7$ GeV be the energy of
a muon giving a 2.6 m track in water.

We find:
\begin{eqnarray}
{\rm FC}(E_\mu,c_\mu) = { 2\over V } \,  
\int_0^R dz\, \int_0^{2\sqrt{R^2-z^2}} dx\,\int_0^H dy\;
\Theta[d(x,y,z)-2\,{\rm m}] \cr
\Biggl\{\Theta\left[{2\sqrt{R^2-z^2}-x}-{y\, |t_\mu|}\right]
\Theta\left[{y }-R_w(E_\mu)\,|c_\mu |\right] \cr 
+~\Theta\left[{2\sqrt{R^2-z^2}-x}-R_w(E_\mu)\,|s_\mu |\right]
\Theta\left[ y\,|t_\mu | +x-{2\sqrt{R^2-z^2}}|\right]\Biggr\}
\label{FC}
\end{eqnarray}
\begin{eqnarray}
{\rm PC}(E_\mu,c_\mu) = { 2\over V } \,  \Theta[R_w(E_\mu)-R_w(E_{min})]\,
\int_0^R dz\, \int_0^{2\sqrt{R^2-z^2}} dx\,\int_0^H dy\;  \cr
\Theta[d(x,y,z)-2\,{\rm m}]  \;
\Biggl\{\Theta\left[{2\sqrt{R^2-z^2}-x}-{y\, |t_\mu|}\right]
\Theta\left[R_w(E_\mu)\,|c_\mu |-y\right] \cr 
+~\Theta\left[R_w(E_\mu)\,|s_\mu |-{2\sqrt{R^2-z^2}+x}\right]
\Theta\left[ y\,|t_\mu | +x-{2\sqrt{R^2-z^2}}|\right]\Biggr\}
\label{PC}
\end{eqnarray}
where the range of muons in water that we use, $R_w(E_\mu)$, 
can be obtained from the expressions in~\cite{PDG}.

A muon produced with energy $E_\mu$, after travelling  a distance $l$
in rock material, has an energy $E'_\mu=R_r^{-1}[R_r(E_\mu)-l]$;
its remaining range in water is $l_w(E_\mu,l)= R_w(E'_\mu)$.
 The effective area for through-going muons of Eq.~(\ref{thru}) is then given by
\begin{eqnarray}
{\rm A}(E_\mu,c_\mu) = { 2\over R_r(E_\mu)-R_r(E'_{min}) } \,  
\int_0^{R_r(E_\mu)-R_r(E'_{min})} dl\, \int_0^R dz\,
  \cr
\Biggl\{|s_\mu |\int_0^H dy \;
\Theta\left[l_w(E_\mu,l)-{\rm Min}\left({2\,\sqrt{R^2-z^2}\over |s_\mu |} ,
{ [H-y]\over |c_\mu |}
\right)\right]\cr
\Theta\left[{\rm Min}\left( {2\,\sqrt{R^2-z^2}\over |s_\mu |} ,
{ [H-y]\over |c_\mu |}\right)-7\,{\rm m} \right] \cr 
+\, |c_\mu |\int_0^{2\,\sqrt{R^2-z^2}} dx \;
\Theta\left[l_w(E_\mu,l)-{\rm Min}\left({2\,\sqrt{R^2-z^2}-x\over |s_\mu |} ,
{ H\over |c_\mu |}
\right)\right]\cr
\Theta\left[{\rm Min}\left( {2\,\sqrt{R^2-z^2}-x\over |s_\mu |} ,
{ H\over |c_\mu |}\right)-7\,{\rm m} \right]\Biggr\},
\label{A}
\end{eqnarray}
where for the range of muons in rock $R_r(E_\mu)$, we have used the results in  \cite{skup}. 

All of the above expressions can be integrated explicitly,
but the analytical results are not brief, or useful.

\newpage

{\Large{\bf Appendix D: Comparison with the SK Monte Carlo}}

\begin{figure}[htb]
  \centering
\mbox{\epsfig{file=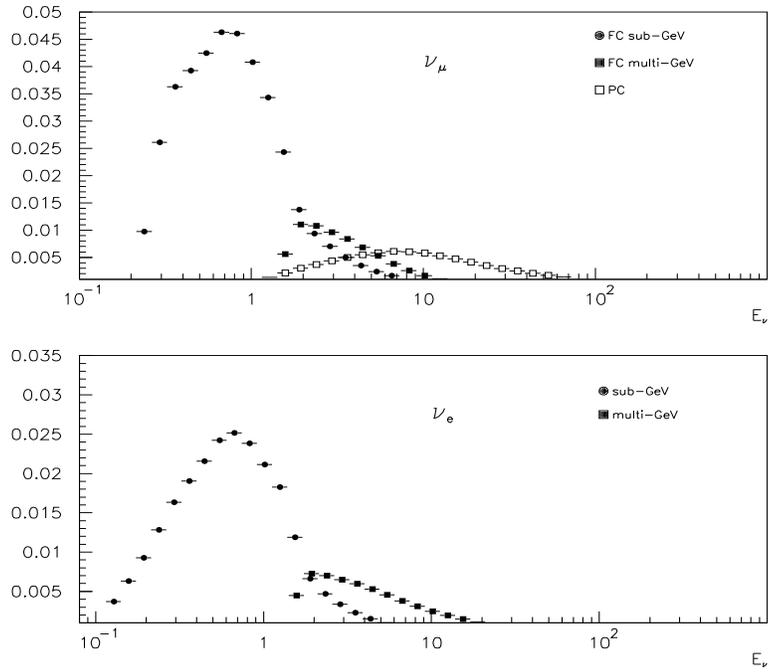,width=4.5in,height=4.in}}
  \caption{Parent $\nu$ energies for $\mu$-like (sgev, FC-mgev, PC-mgev) and 
$e$-like (sgev and mgev) events.}
\label{parent}
\end{figure}

\begin{figure}[htb]
  \centering
\mbox{\epsfig{file=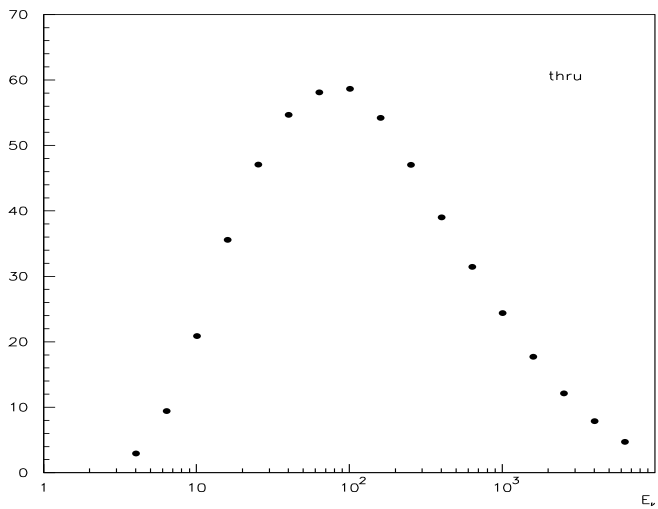,width=4.in,height=3.in}}
  \caption{Parent $\nu$ energies for through-going muons.}
\label{parentup}
\end{figure}

We have not included in this analysis any non-geometrical 
detection efficiencies, as discussed in Section 3.
We have normalized the number of events
in each data sample to the corresponding number in the SK Monte Carlo for the 
non-oscillation hypothesis.
This is tantamount to the use of an
efficiency function which is not a function of energy
within each sample.
 A non-trivial check that this is indeed a sensible approximation is to
 compare 
the parent neutrino energy distributions in the different data samples with those
worked out by the SK team. 
These distributions are defined as the azimuthally averaged neutrino flux weighted with the 
integrated neutrino cross section and with the selection function for the 
various
sgev and mgev data samples:
\begin{eqnarray}
 P_{s,l}(E_\nu) \propto  \int   dc_\nu\, dE_l\, 
dc_\beta\, d\phi \; {\rm Th_{s,l}}(E_l,c_l) \; 
  \;  \left[ \sigma(E_\nu, E_l, c_\beta)\, {d\Phi_{\nu}\over dE_\nu dc_\nu} +
{\bar \sigma}(E_\nu, E_l, c_\beta)\, 
\frac{d{\bar\Phi}_{\nu}}{dE_\nu dc_\nu}\right]\, . 
\nonumber
\end{eqnarray}
All of the symbols in this expression have already been defined.
Our results, shown   
in Figs.~\ref{parent} and \ref{parentup}, 
are in good agreement with those obtained by SK \cite{kate2}. 

\begin{figure}[htb]
  \centering
\mbox{\epsfig{file=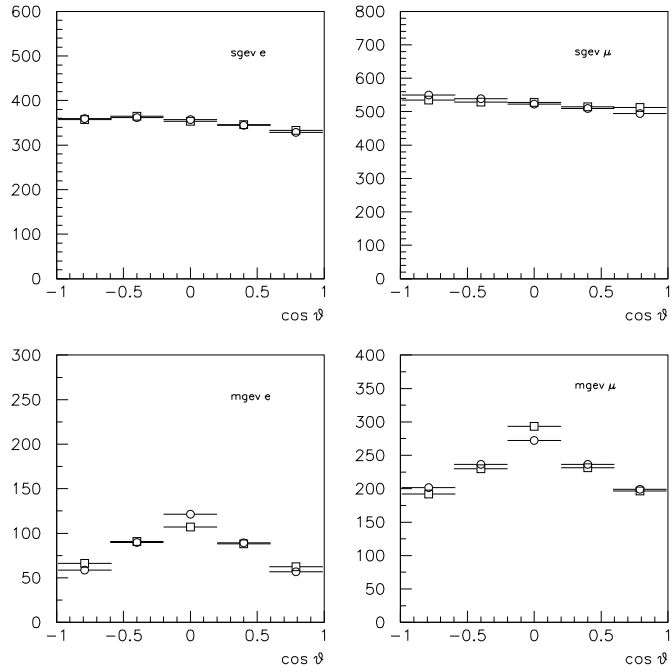,width=4.in,height=4.in}}
  \caption{Zenith-angle distributions of $e,\mu$ sgev and mgev samples. 
The squares are the SK Monte Carlo results and the circles are our
predictions, both for the no-oscillation hypothesis. The total areas under 
the curves are normalized to be the same. }
\label{zenithno}
\end{figure}

\begin{figure}[htb]
  \centering
\mbox{\epsfig{file=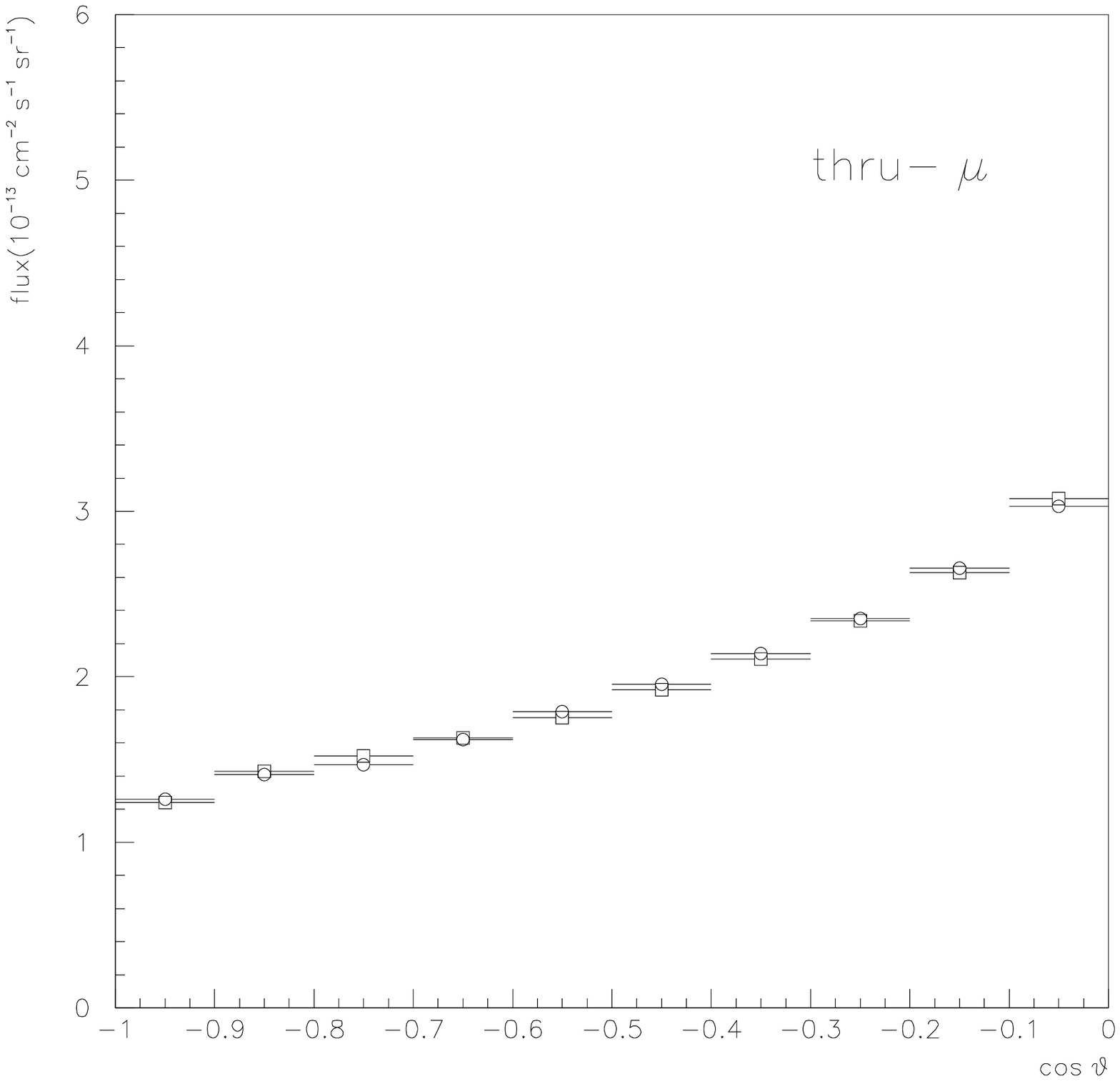,width=4.5in,height=3.in}}
  \caption{Zenith-angle distributions of through-going muons for the 
non-oscillation hypothesis. The
squares are SK Monte Carlo results and the circles are our predictions.}
\label{zenithup}
\end{figure}

In Figs.~\ref{zenithno} and \ref{zenithup} we compare the zenith angle dependence obtained in our calculation for the non-oscillation hypothesis with the predictions of 
the SK Monte Carlo \cite{hayato}. The agreement is again rather good.

\begin{thebibliography}{9}
\bibitem{kate} We warmly thank K. Scholberg for providing us with the official SK
data, also shown in various talks: see for instance \cite{hayato}.
\end{thebibliography}
\end{document}